# Revolutionary valorization of carbon dioxide into dimethyl carbonate is catalyzed by sodium chloride: cheap, clean, one-pot, and water-free synthesis

Vitaly V. Chaban,[1] Nadezhda A. Andreeva,[2] Leonardo Moreira dos Santos,[3] and Sandra Einloft[3,4]

(1) Yerevan State University, Yerevan, 0025, Armenia. E-mail: vvchaban@gmail.com.
(2) Peter the Great St. Petersburg Polytechnic University, Saint Petersburg, Russian Federation.
(3) School of Technology, Pontifical Catholic University of Rio Grande do Sul - PUCRS, Porto Alegre Brazil.
(4) Post-Graduation Program in Materials Engineering and Technology, School of Technology Pontifical Catholic University of Rio Grande do Sul – PUCRS, Brazil.

**Abstract:**

The robust valorization of carbon dioxide ($CO_2$) stays at the center of sustainable development. Since $CO_2$ represents a low-energy compound, its transformation into commercially coveted products is cumbersome. In the present work, we report a revolutionary method to obtain dimethyl carbonate (DMC) out of methanol ($CH_3OH$) and $CO_2$ catalyzed by sodium chloride (NaCl) and similar inorganic salts. The computational exploration revealed a mechanism of favorable catalysis, which was subsequently confirmed experimentally. Unlike all competitive syntheses of DMC, the new one does not produce water and, therefore, the hydrolysis of a carbonate does not occur. No dehydrating agents are necessary. The employed catalyst is cheap and permanently exists in the same phase with the reactants and products. The action of NaCl was compared to those of other alkali metal salts, LiI, LiCl, and KI, and competitive performances were recorded. The experimentally obtained result outperforms most competing technologies according to the DMC yield, 19% with molecular sieves and 17% without molecular sieves. All existing competitors are excelled by the simplicity and cleanness of the synthesis. The reported advance substantially simplifies the synthesis of linear organic carbonates and robustly valorizes CO$_2$.

**TOC Image**

Sodium chloride makes the production of dimethyl carbonate way more straightforward.

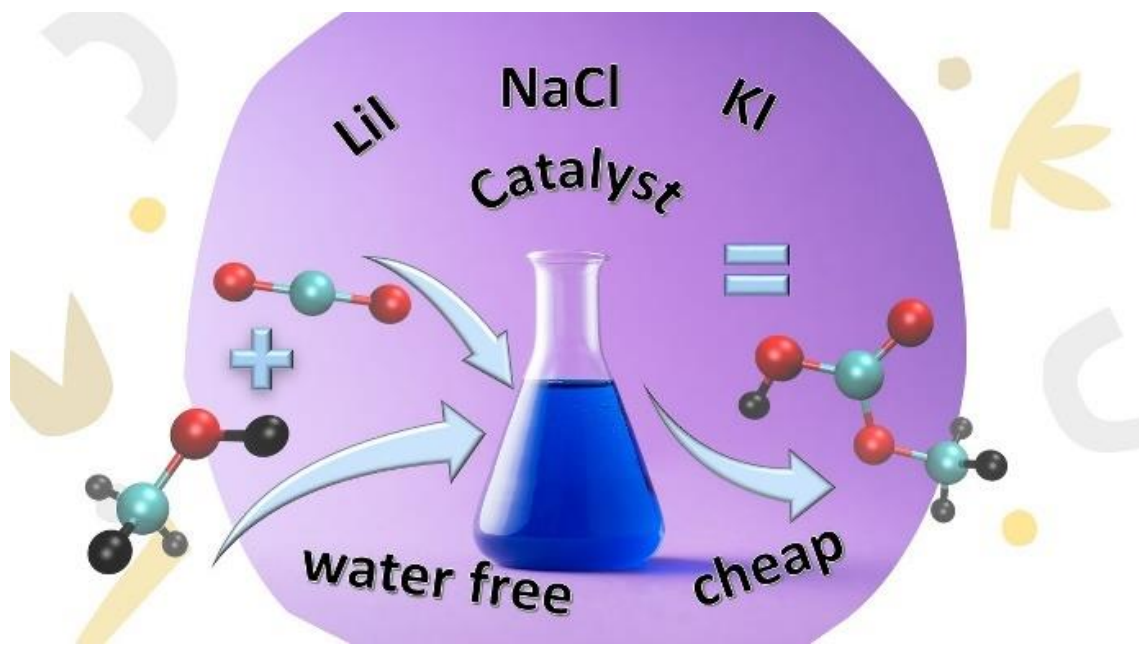

## 1. Introduction

The extra concentration of carbon dioxide ($CO_2$) in the atmosphere of the Earth is seen as a factor leading to emerging global warming.[1] The peak average yearly temperatures were recorded during the last several years. Furthermore, the steady large-scale temperature increase, allegedly, happens for more than a century. The time frame somehow coincides with the bost of large-scale chemical technologies. These global changes reflect undesirable climate modifications, the thawing of glaciers, unpredictable and unstable evolutions of the existing ecosystems. The sustainable development of humanity implies that current $CO_2$ concentration is monitored and on-time measures are taken to stabilize it.[1] More and more advanced fixation and conversion of greenhouse gases is pursued worldwide[2-6] by adapting novel materials and approaches.[7-10]

$CO_2$ represents one of the most thermodynamically stable molecules composed of carbon and oxygen. Therefore, it appears to be chemically challenging to design novel energetically favorable reactions involving $CO_2$ as a reactant. In the meantime, $CO_2$ is not just an environmental hazard for people but also a valuable source of carbon for building live creatures and accumulating energy, e.g., in the form of starch. Live nature organisms, such as plants and bacteria, learned to rather efficiently apply $CO_2$ in photosynthesis to produce carbohydrates. Furthermore, certain bacteria elaborated the noteworthy chemical pathways to valorize this greenhouse gas into methanol ($CH_3OH$) and other small organic molecules. Teaching microbes to produce aqueous ethanol solutions out of $CO_2$ is definitely the next worthwhile educational challenge for microbiologists. The already-known versatility of the sophisticated $CO_2$ chemistry around us is noteworthy, notwithstanding its obvious thermodynamic stability.[11] Similar to solar energy, $CO_2$ is an abundant and virtually free resource for our civilization. The robust valorization of $CO_2$ represents the means to create a more sustainable economy and mitigate adverse changes in the planetarian climate.

The two major sources of greenhouse gases are the burning of fossil fuels (mainly $CO_2$) and agriculture (mainly $CH_4$). Animals persistently exhale $CO_2$. Finding routes to either employ $CO_2$ and $CH_4$ directly or synthesize valuable products out of them would be an essential step toward sustainability. In pure state, $CO_2$ can be used as a refrigerant, propellant, and fire extinguisher. In chemical technologies, $CO_2$ is a reactant to produce fuels, plastics, and task-specific chemicals, such as alcohols, carboxylic acids, dialkyl, and cyclic organic carbonates. The industry of $CO_2$ valorization remains in its early stages. As it develops, new jobs in chemical and manufacturing industries will emerge and innovative products may be proposed. The perspective of using $CO_2$ as a feedstock for new products deserves a profound investigation nowadays.[12]

Organic and inorganic carbonates are genetically close to $CO_2$.[13-16] Carbonates can be arguably seen as condensed/trapped $CO_2$.[17] Mineralization designates a process, in which carbonate-containing minerals are derived out of $CO_2$. These products are versatile construction materials.[18-20] In the form of inorganic carbonates, the captured volumes of $CO_2$ can be stored for, at least, centuries. In turn, organic linear and cyclic carbonates can be obtained via the reactions of $CO_2$ with alcohols or epoxides. Organic carbonates are important solvents for applications.[21-26] They are conventionally used in mixtures with more polar solvents to tune the dielectric constants of the liquid media.[27-30] Whatever technology is chosen for $CO_2$ valorization, it is of paramount importance that the ultimate products be commercially competitive. This criterion means that all stages of the chemical transformations responsible for valorization must demand adequate amounts of energy, catalysts, and co-reactants.

In the present work, we boost the valorization of $CO_2$ to obtain dimethyl carbonate (DMC). $CH_3OH$ acts as a co-reactant in this synthesis. Whereas this is a vivid field of research, the achieved progress has been fairly modest thus far.[34-42] The developed catalysts partly eliminate the high activation barrier to generate methoxy moieties by deprotonating $CH_3OH$ molecules. Furthermore, the thermochemistry of the entire reaction is generally unfavorable.[14,31-33] As a result, most published works report the DMC yields of 5-15 mol%, notwithstanding extreme conditions of the synthesis (up to 180 degrees Celsius and up to 300 bar) and the need to dehydrate samples.[18,31,34-36]

d-Metal oxides distributed over suitable inorganic matrices are presently used to abstract the proton from $CH_3OH$.[31] For instance, $Fe_3O_4$, $CeO_2$, $ZnO$, $ZrO_2$, $TiO_2$, $La_2O_3$, $MgO$, $Al_2O_3$, $SnO_2$, $CuO$, $CuNi$, $TiO_2$, $FeO$, and numerous noteworthy catalytic setups[14,36-44] were employed under various reaction conditions to enhance the selectivities and conversions of $CH_3OH$. As a rule, more chemically active metals exhibit better performances but morphologies of the resulting catalysts and matrices are also very essential. The metal atom coordinates the hydroxyl group of $CH_3OH$ and polarizes the oxygen-hydrogen bond. The oxygen atom of the catalyst does the same with the hydroxyl hydrogen. The elongated covalent bond in hydroxyl helps the proton to dissociate. The proton abstraction represents the rate- and yield-limiting stage of the reaction. Another problem of the existing approaches is the formation of water, which hydrolyzes DMC.[45] The dehydrating agents are employed to combat the reverse reaction. Such a circumvention improves the yields but essentially complicates the experimental setups making the results hard to interpret and rationalize.[45]

Herein, we abandoned metal oxides and trialed alkali metal salts – NaCl, KI, LiI – as catalysts. Unlike competitors, the latter are soluble in $CH_3OH$. The cations coordinate $CH_3OH$ molecules strongly. Thus, the total surface of contact between the catalyst and the reactants is maximized and their usage gets more efficient. $CH_3OH$ can be supplied in liquid or gaseous states. $CO_2$ is permanently in the gaseous state. The elevated pressure accelerates the reaction and improves the thermochemistry thanks to suppressing the entropic penalty. To initiate the conversion of $CH_3OH$, the methoxy group, $CH_3O^*$, must first be formed. The role of the dissolved inorganic ions is to stretch the hydroxyl moiety of $CH_3OH$, thus promoting hydrogen abstraction upon thermal motion. The inorganic anion (chloride, iodide) thermodynamically stabilizes the abstracted proton. Simultaneously, $CH_3O^*$ readily forms the covalent bond with $CO_2$, which transforms into the carboxyl moiety, giving $CH_3OC(O)O^-$. The alkali cation or the abstracted proton temporarily coordinates the new anion due to the Coulombic attraction. The reaction is finalized via the attachment of the methyl group, which comes from the promoter molecule, $CH_3OC(O)O^- + CH_3I = DMC + I^-$. The liberated iodide stabilizes the excess proton. Most importantly, the above-described route does not involve water, the most undesirable by-product in all competitor formulations. The computational investigation has been concentrated on the most energetically demanding transformation, whereas the synthesis has included all essential experimental descriptors of the aggregate reaction and verified the theoretical predictions.

## 2. Computational methodology

The simulated systems were composed of three $CH_3OH$ molecules, one of which is a reactant and two of which are solvents. The catalyst was represented as a single ion pair (NaCl, LiI, and KI). One $CO_2$ molecule was located nearby. The described chemical compositions were immersed into a polarizable continuum with the physicochemical properties of $CH_3OH$.

The work makes use of numerous computational methods, such as local optimization to the closest stationary point (local minimum and transition state),[46-48] potential energy landscape exploration along the chosen reaction coordinates, global minimum configuration search, single-point calculations using semiempirical[49-50] and hybrid density functional theory (HDFT) Hamiltonians,[51] Born-Oppenheimer PM7-MD simulations,[52-55] thermochemistry, and reaction activation barrier evaluations. The following paragraphs characterize the parameters of the methods, that were employed, and argue the methodological choices.

The reported geometries and electronic properties of the reactants, products, and transition states were obtained in the framework of HDFT. The exchange-correlation, range-separated, meta-general-gradient-approximation functional M11 was employed to obtain electronic structures.[51] The self-consistent-field convergence criterion of $10^{-8}$ hartree was applied. The empirical correction to account for the longer-range attractive forces (dispersion) acting among non-covalently bound atoms was applied on the fly. The split-valence triple-zeta polarized basis set TZVP was used[56] to supply elementary functions describing electronic orbitals of chemical elements, out of which the molecular wave function was constructed for each simulated system.

The PM7 Hamiltonian was employed to conduct a global potential energy minimum search for the structures of chemical interest (reactants and products) and molecular dynamics simulations.[49-50] The latter were used to walk across activation barriers and record the conformational versatility of the simulated systems. The wave function convergence criteria of $10^{-4}$ kcal/mol were employed in the self-consistent field algorithm.

The local optimizations of the molecular geometries were conducted through the rational function optimization algorithm. The simultaneously required convergence criteria for a stationary point were set as follows: 0.6 kJ pm$^{-1}$ for the strongest nuclear force, 0.04 kJ nm$^{-1}$ for the RMS nuclear force, 0.9 pm for the largest nuclear displacement, and 0.6 pm for the RMS nuclear displacement. The procedure operated Cartesian coordinates of all atoms.

The chosen reaction coordinates were scanned stepwise in the vicinity of the anticipated transition states. While the reaction coordinate remained rigidly fixed at a certain value, all other degrees of freedom were free to evolve based on the energy gradients. The restrained local geometry optimization was carried out at every value of the reaction coordinate to find the minimum energy. Each reaction coordinate was propagated with a linear step of 1-2 pm. The calculations of the vibrational frequency profiles were conducted for maxima points along the obtained reaction profiles. The frequencies were derived using the conventional rigid rotor approximation. The heights of the maxima relative to the reactant energies represent activation barriers for the considered chemical transformation. All wave functions related to the reaction profiles were computed by the unrestricted version of the HDFT calculations.

The global minimum energy stationary points were found via the kinetic energy injection method described elsewhere.[46-48] One hundred candidate structures per chemical composition were considered. The perturbation kinetic energy was set to 1000 K. The excess energy was injected into systems every 0.1 ps. PM7-MD simulations were used to propagate the nuclear trajectories.

The thermochemical potentials were derived from molecular partition functions in the implicit solvent methanol. The molecules at each state (reactants and products) were treated jointly. The gas phase approximation was not used. For $CO_2$, it meant that its dissolved state in liquid $CH_3OH$ instead of the gas state was considered. The sequence from Hess's law was applied to get the potentials (Gibbs free energy, enthalpy, entropy) of the chemical reactions. The obtained frequencies were used to compute near- and far-infrared spectra and obtain zero-point energy corrections. It has been added to electronic energies.

The polarizable continuum approach was employed to model the implicit solvation in $CH_3OH$. The model performs by modifying the Hamiltonian of the system with a solute-solvent coupling term depending on the solute's local electric field.[57] The effect of solvation was included in all reported properties.

The electronic polarization was described via partial electrostatic charges obtained according to the Merz-Singh-Kollman algorithm.[58] Throughout the discussion, we use partial atomic charges as quantitative measures of nucleophilicities and electrophilicities of the reactants' interaction centers.

The charge transfer amounts were assessed via the natural bond orbital (NBO) analysis. These processes were characterized by the interaction energies between the localized orbitals. The energies were computed according to the second-order perturbation theory. Furthermore, the NBO charges on atoms were computed to compare the properties of various interaction sites.

Gaussian09 software was applied to conduct electronic-structure calculations.[57] VMD 1.9.3[59] and Gabedit 2.5.2[60] were used to manipulate the Cartesian and internal coordinates of the simulated systems. The in-home software was used to conduct non-standard simulations and process the numerical results.

## 3. Experimental methodology

### 3.1. Materials

Methanol (>99,9% -EMSURE®), sodium chloride ( ≥ 99,5%, EMSURE®), iodomethane (>99.5%, ALDRICH), diethyl ether (>99.9%, EMSURE®), DMC (>99.5%, ALDRICH), pearl-shaped molecular sieves (3A, ALDRICH), and $CO_2$ (99,8%, White Martins).

### 3.2. DMC synthesis

The reactions were conducted in the 120 ml metal reactor with constant magnetic stirring. The temperature was controlled using a thermocouple connected to the temperature controller along with a heating mantle around the reactor. In reactions using the dehydrating agent, the reactor was equipped with an upper metallic compartment (gas phase) filled with molecular sieves. The reactions employed 213 mmol of $CH_3OH$, and different amounts of NaCl (5.134, 8.556, 11.98, 15.40, 18.82, 25.67, and 34.22 mmol NaCl), and 20 mmol of $CH_3I$. The reactions were carried out at 40 bar of $CO_2$ pressure and a temperature of 80ºC. Each reaction proceeded 24 hours before the analysis of the products was performed.

All tests were performed in triplicates. The products from the catalytic tests were analyzed by gas chromatography (GC) to determine yield, conversion, and selectivity. The Shimadzu GC-2014 gas chromatograph equipped with an SH-Rtx-5 column was used. The temperature schedule involved a heating ramp of 31°C for 0.5 minutes, 10°C min$^{-1}$ to 50°C for 1 minute, 20°C min$^{-1}$ to 100°C for 2 minutes, and 50°C min$^{-1}$ to 220°C for 2 minutes. The samples were diluted with a 4% (v/v) concentration in ethyl ether. The area of the DMC peak (2.4-2.7 minutes) was used to determine the concentration through the pure DMC calibration curve, following procedures described by Valente, Riedo, and Augusto (2003).[61] The conversions and selectivities were calculated as described elsewhere by Faria and coworkers.[14,31,62] The pH measurements were carried out using the Digimed DM20 pH meter.

## 4. Results and discussion

The theoretical part of this work deals with the most energetically demanding stage of the DMC synthesis, which is proton abstraction, i.e., the formation of CH$_3$O*. Once formed, the metholate radical readily captures CO$_2$. We provide a comprehensive description of the reactants, transition states, and products involved in this cornerstone process for DMC synthesis. We deliberately used lone cations and ion pairs as separate catalysts to understand how the concentration of the inorganic salt in CH$_3$OH influences the catalytic performance. The lone ions prevail in dilute solutions, whereas ion pairs are abundant in more concentrated solutions. By comparing both mentioned options, the computer simulations may suggest optimal experimental setups.

### 4.1. Global minimum search

It was of essential importance to identify the most thermodynamically stable geometry of the intermediate. The $CH_3OC(O)O^-$ anion could be neutralized by $H^+$, which leaves the reacting $CH_3OH$ molecule, or the alkali metal cation ($Li^+$, $Na^+$, $K^+$). Furthermore, the resulting structure could be solvated in different ways thanks to a few hydrogen bonding sites available. The global minimum search revealed about ten local minimum conformations of every composition. Figure 1 exemplifies the obtained potential energies for the systems containing LiI as a catalyst, whereas other compositions exhibit qualitatively similar behaviors. The versatilities of the obtained structures occur due to the various possible arrangements of H-bonds and the competition of the proton with alkali cations for the location near the oxygen atoms of the hydroxyl groups within the reactants and near the carboxyl group within the products.

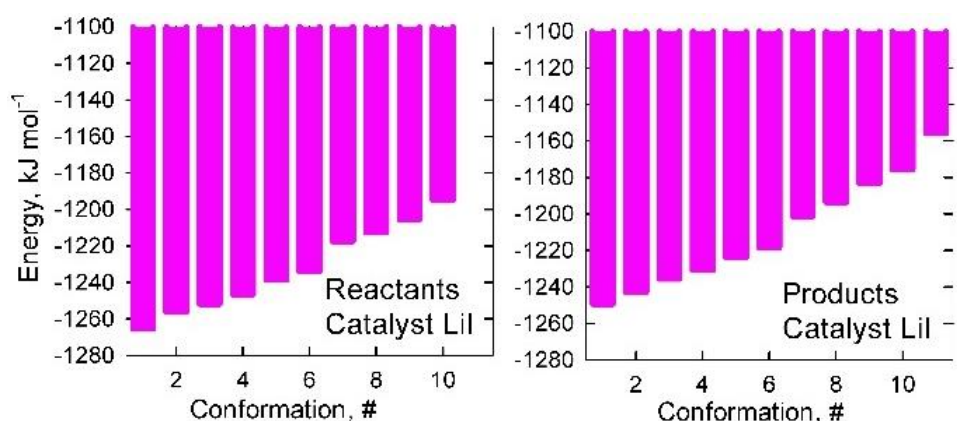

Figure 1. The potential energies of the stationary points in the reactants (3 $CH_3OH$, 1 $CO_2$, and LiI catalyst) and products (1 $CH_3OC(O)O^-H^+$, 2 $CH_3OH$, and LiI catalyst).

Figure 2 visualizes the selected global-minimum structures containing $Li^+$ and LiI as catalysts. The alkali cation strongly coordinates a few oxygen atoms of $CH_3OH$ and the emerged carbonate moiety. Another strong attraction takes place between the former hydrogen atom of $CH_3OH$ and the iodide anion. The oxygen and iodine atoms compete for the proton, which is located between them. The two stationary points of similar potential energy stand for the proton covalently attached to iodine and H-bonded to oxygen and vice versa. By definition, iodine does not participate in H-bonding. Yet, its electrostatic attraction to hydrogen is strong. In the systems without the anion (chloride, iodide), the hydrogen atom is bound to the carbonate moiety in the corresponding global minimum states but may be electrostatically attracted to the solvent ($CH_3OH$) molecules in higher-energy stationary states.

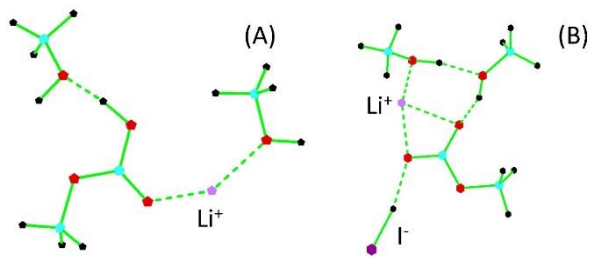

Figure 2. The selected low-energy molecular configurations of the products: (A) catalyst $Li^+$ and (B) catalyst catalyst LiI. The dashed lines highlight strong electrostatic interactions, including H-bonds.

Figure 3 compares the molecular geometries of the reactant and product compositions, whereas the reaction is catalyzed by the potassium cation. The cation coordinates three $CH_3OH$ molecules and the $CO_2$ molecule. The hydroxyl groups are polarized thanks to such an electrostatic attraction. In turn, after the reaction, $K^+$ coordinates only one $CH_3OH$ molecule and the newly formed carbonate group. The remaining solvent molecule diffuses away from the catalyst and forms a strong H-bond, 170 pm long, with the protonated carboxyl group of the product.

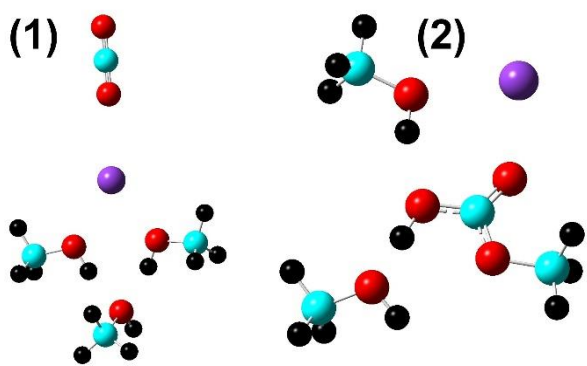

Figure 3. The optimized molecular structures were obtained for (1) reactants and (2) products. The employed catalyst is the potassium cation.

If $K^+$ is provided as an ion pair, KI, its coordination capability essentially deteriorates (Figure 4). However, the payoff is that the iodide anion coordinates the hydrogen atom of $CH_3OH$.

As we rationalize below in terms of thermochemistry and activation barriers, this feature plays a paramount role in the simpler deprotonation of the hydroxyl moiety. In the state of products, the location of the catalyst does not change substantially. The reacted system remains to be strongly correlated via the H-bonded network.

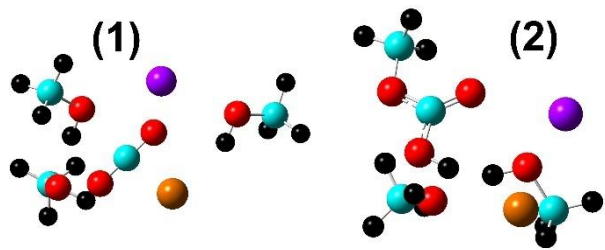

Figure 4. The optimized molecular structures were obtained for (1) reactants and (2) products. The employed catalyst is potassium iodide.

Finding global minimum states for the involved stationary points (reactant, transition state, reaction intermediate, product) is a cornerstone of a trustworthy energetic evaluation. The height of the activation barrier represents an energetic difference between the transition state geometry and reactant geometry. Therefore, choosing a non-global minimum configuration of the reactants proportionally underestimates the applicable activation barrier. The same philosophy should apply to the transition states. While a few transition states may correspond to a given chemical reaction, only the energetically lowest one is followed in real-world syntheses. This lowest activation barrier unequivocally determined the cheapest reaction pathway.

### 4.2. Binding and cohesion energies

The non-covalent binding energies (Figure 5) within the simulated systems reflect the relative conformational rigidities. On a related note, strong interactions within reactants and products, as well as the solvation energy gain, thermodynamically stabilize the respective system. Consequently, the non-covalent interactions contribute to the thermochemistry of the studied reactions. The strongest interactions take place within the products upon using the NaCl catalyst, the corresponding potential energy amounting to -798 kJ/mol. In turn, the corresponding reactants exhibit a total non-covalent binding energy of -796 kJ/mol. The difference with other systems comes from the strong hydrogen-chloride H-bonding and strong sodium-oxygen electrostatic attraction. Furthermore, weaker attractions exist between the alkali metal cation and the three remaining polar oxygen atoms located in the proximity. The setups containing the chloride anion appear more correlated with one another as compared to the iodide-containing systems. The present observation suggests that chloride is a somewhat more desirable counterpart of the catalyst. Large negative binding energies in the LiI-catalyzed samples are likely due to the strong coupling between the smallest possible alkali cation with the adjacent nucleophilic centers. Such a feature of lithium is routinely employed in chemical energy sources.[63-64]

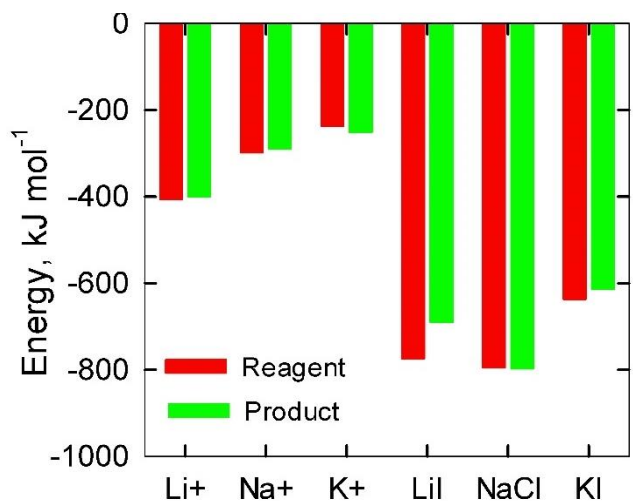

Figure 5. The sums of the non-covalent binding energies in the reactants (red bars) and products (green bars).

The simulated systems were immersed into the implicit solvent, $CH_3OH$. Table 1 summarizes the molar volumes depending on the molecular compositions of the systems. The $CO_2$ capture reaction increases the molar volume in all cases except for the KI catalyst. Bulky $K^+I^-$ is known to exist as a rather diffuse ion pair.[65] The formation of the products fosters a more energetically favorable reorientation of these ions. The latter results in a more compact structure, which exhibits a somewhat lower molar volume. Note that the reactant configurations may be designated as $CO_2$ physisorbed in $CH_3OH$. In turn, the product configurations may be designated as chemisorbed $CO_2$.

Table 1. The molar volumes of the systems containing physisorbed (reactants) and chemisorbed (products) $CO_2$.

| Process | $Li^+$, $cm^3/mol$ | $Na^+$, $cm^3/mol$ | $K^+$, $cm^3/mol$ | LiI, $cm^3/mol$ | NaCl, $cm^3/mol$ | KI, $cm^3/mol$ |
|---|---|---|---|---|---|---|
| Reactants | 88.3 | 87.5 | 108 | 123 | 145 | 168 |
| Products | 113 | 99.3 | 114 | 153 | 187 | 141 |

The cohesion energies (Figure 6) support the above conclusion regarding the strongest physical interactions in the NaCl-containing system. Furthermore, one notices that the ion pair catalysts exhibit stronger cohesions than the lone cation catalysts. The effect of the anion was not evident without specific calculations. For instance, one could hypothesize that an anion could screen the electrostatic interactions between the cation and the hydroxyl group and, therefore, quench the polarization of the proton originally belonging to the reacting $CH_3OH$ molecule. The simulations revealed that the anions play an essential role in coordinating the hydrogen atom being favorable for deprotonation.

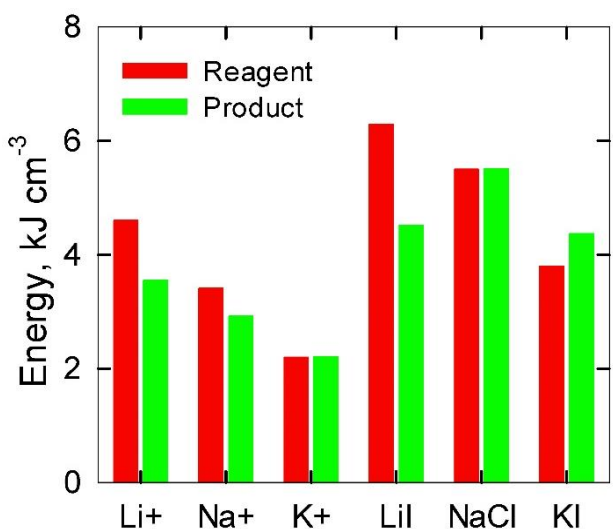

Figure 6. The cohesion energies in the simulated systems represent physisorbed $CO_2$ (red bars) and chemisorbed $CO_2$ (green bars).

**4.3. Bond lengths, atomic nucleophilicities, and atom-atomic charge transfers**

In a strongly correlated system, numerous electron transfers can be identified occurring between higher- and lower-energy orbitals. The most essential charge transfers involve the lone electron pairs of strongly electronegative interaction centers and empty non-bonded orbitals of the more electropositive interaction centers. Such processes decrease the strength of the covalent bonding, if any exists, and increase physical interactions. Specifically, the electron transfers between orbitals are responsible for the strong solvation of small cations in their solutions.

The oxygen atoms of $CH_3OH$ possess lone electron pairs, which interact with the alkali metal cations. The corresponding energies in the case of lithium amount to 96 and 88 kJ/mol. These interactions indicate very strong atom-atomic couplings. Furthermore, $Li^+$ interacts with one of the oxygen atoms of $CO_2$ with an energy of 79 kJ/mol. The presence of the iodine anion substantially quenches the mentioned interactions. The corresponding energies equal 65 kJ/mol with $CH_3OH$ and 11 kJ/mol with $CO_2$. Noteworthy, no noticeable charge transfer takes place between iodine and hydrogen even though these atoms interact strongly based on the optimized molecular geometries. Their attraction is to a large extent due to Coulombic forces and to a lesser extent due to the overlap of electronic orbitals. For instance, the hydrogen-iodine distance in the KI-catalyzed system amounts to 250 pm.

Table 2 summarizes the strongest physical interactions in the simulated systems. These interactions undermine the stabilities of the covalent bonds and contribute the equivalent portions of potential energy to the stabilization of ion-ionic and ion-molecular couplings. All the interactions take place among the most electronegative sites, oxygen atoms, and three alkali metal cations. The oxygen atoms may belong to any of the $CH_3OH$, $CO_2$, and carboxyl moieties. To exhibit a strong mutual attraction following the charge transfer mechanism, the atoms must be located closer than 300 pm. It means that smaller alkali cations exhibit stronger non-covalent attractions to the oxygen atoms. Compare, a very strong interaction involving $CH_3OH$ and $Li^+$ of 96 kJ/mol, whereas the optimized distance between them amounts to 197 pm (Table 3). Interestingly, the interaction of $Li^+$ with the protonated carboxyl group is 59 kJ/mol at a distance of 198 pm. The protonation state definitely impacts the amount of charge transfer.

Table 2. The strongest stabilization energies emerging due to partial charge transfers were calculated via the NBO-based analysis. LP designates a lone electron pair. LP* designates an unfilled valence-shell non-bonding orbital.

| Pairwise charge transfer | Energy, kJ mol$^{-1}$ | | | | | |
|---|---|---|---|---|---|---|
| | $Li^+$ | $Na^+$ | $K^+$ | LiI | NaCl | KI |
| Reactants | | | | | | |
| LP($O^{CO2}$)→LP*(cation) | 79 | 19 | 26 | 11 | 57 | 52 |
| LP($O^{CH3OH}$)→LP*(cation) | 96 | 28 | 22 | 65 | 67 | 47 |
| Products | | | | | | |
| LP($O^{carboxyl}$)→LP*(cation) | 59 | 27 | 17 | 57 | 46 | 51 |
| LP($O^{CH3OH}$)→LP*(cation) | 52 | 24 | 12 | 59 | 42 | 40 |

An important feature of the $CH_3OH$-inorganic salt systems is strong ion-molecular coupling. For instance, all cations strongly attract the $CO_2$ molecule. The charge-transfer-based attraction between $K^+$ and $CO_2$ equals 52 kJ/mol before the reaction. The distance from $K^+$ to one of the oxygen atoms of $CO_2$ is 304 pm. The systems containing both cation and anion generally exhibit large charge transfers as compared to the systems with a cation only. The revealed interactions are commensurate with the high cohesion energies discussed above. We conclude that the alkali cation polarizes not only the $CH_3OH$ but also the $CO_2$ molecules, which it coordinates.

Table 3. The non-covalent distances, in picometers, between the oxygen atoms and the alkali metal cations in the investigated systems.

| Interaction | $Li^+$ | $Na^+$ | $K^+$ | LiI | NaCl | KI |
|---|---|---|---|---|---|---|
| Reactants | | | | | | |
| $O^{CO2}$ - cation | 221 | 278 | 293 | 327 | 253 | 304 |
| $O^{CH3OH}$ - cation | 197 | 234 | 268 | 192 | 239 | 276 |
| Products | | | | | | |
| $O^{carboxyl}$ - cation | 198 | 234 | 270 | 202 | 243 | 269 |
| $O^{CH3OH}$ - cation | 196 | 230 | 267 | 194 | 241 | 271 |

The analysis of the NBO charges suggests that the cations and anions possess electron densities close to unity. The presence of the anions in the systems somewhat decreases the charges of the cations. The charges in the product configurations are slightly larger than those in the reactant configurations. The NBO charges belonging to the oxygen atoms, which represent electron donors in the considered interactions, appear substantially lower within products. This feature is in line with our expectations because $CO_2$ acquires an additional electron upon grafting to the methoxy moiety. The obtained electron originally belonged to the hydrogen atom of $CH_3OH$.

Table 4. The partial atomic charges, in atomic units, on the oxygen and alkali metal ions were recomputed from the localized natural orbitals. The derived charges are in quantitative agreement with the other NBO-based properties.

| Species | $Li^+$ | $Na^+$ | $K^+$ | LiI | NaCl | KI |
|---|---|---|---|---|---|---|
| Reactant | | | | | | |
| Cation | +0.90 | +0.96 | +0.96 | +0.89 | +0.87 | +0.86 |
| Anion | — | — | — | -0.91 | -0.89 | -0.90 |
| $O^{CO2}$ | -0.54 | -0.53 | -0.53 | -0.52 | -0.54 | -0.53 |
| $O^{CH3OH}$ | -0.79 | -0.78 | -0.78 | -0.82 | -0.77 | -0.75 |
| Product | | | | | | |
| Cation | +0.95 | +0.97 | +0.98 | +0.95 | +0.91 | +0.89 |
| Anion | — | — | — | -0.93 | -0.90 | -0.88 |
| $O^{carboxyl}$ | -0.73 | -0.70 | -0.53 | -0.75 | -0.70 | -0.68 |
| $O^{CH3OH}$ | -0.73 | -0.79 | -0.78 | -0.80 | -0.76 | -0.78 |

The electrostatic partial atomic charges (Figures 7-8) reflect the electrostatic portion of the non-covalent interactions in the simulated systems representing both reactants and products. The charges are obtained via the ab initio-derived electrostatic potential fitting around the system. The electrostatic charges are frequently used to parameterize the Coulombic fractions of the molecular mechanics Hamiltonians to simulate large atomic ensembles. The charges of the alkali metal cations (Figure 7) are particularly important due to their participation in polarizing the hydroxyl group of $CH_3OH$. Within products, $Li^+$, $Na^+$, and $K^+$ strive to coordinate the newly emerged carbonate moiety as a whole. The carbonate-cation interactions are seen as an important factor that stabilizes the methyl hydrogen carbonate intermediate thermodynamically.

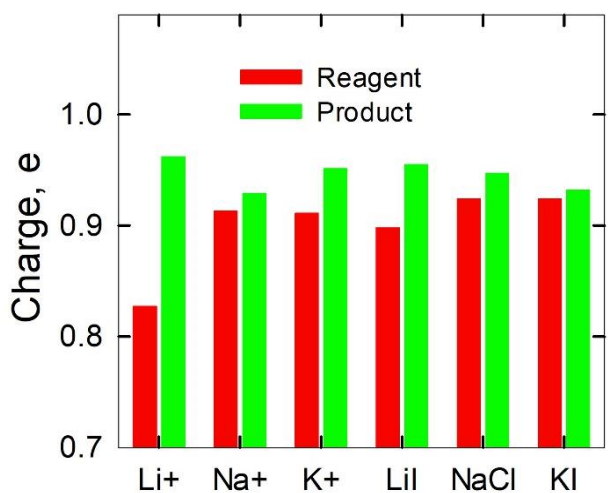

Figure 7. The electrostatic partial charges of the alkali metal cations within the reactants (red bars) and products (green bars) depend on the employed catalyst.

The electrostatic attractions recorded within the products are systematically stronger as compared to those within the reactants. For instance, the charge of $Li^+$ amounts to +0.83e while coordinating the hydroxyl group of the reactant. In turn, the charge of this cation increases up to +0.96e while it coordinates the protonated carbonate moiety. Furthermore, higher partial electrostatic charges imply stronger attractions with the implicitly and explicitly simulated $CH_3OH$ molecules. These interactions are important product-stabilizing factors that enhance the thermochemistry of the considered reaction stage.

The major non-covalent interactions in the simulated systems take place among the oxygen atoms of $CH_3OH$, $CH_3OC(O)OH$, and alkali metal cations. The strengths of these stabilizing interactions depend equally on the cation's charge and the oxygen atoms' charges. Figure 8 compares the charges of the oxygen atoms belonging to the hydroxyl groups of the $CH_3OH$ molecules within reactants and products.

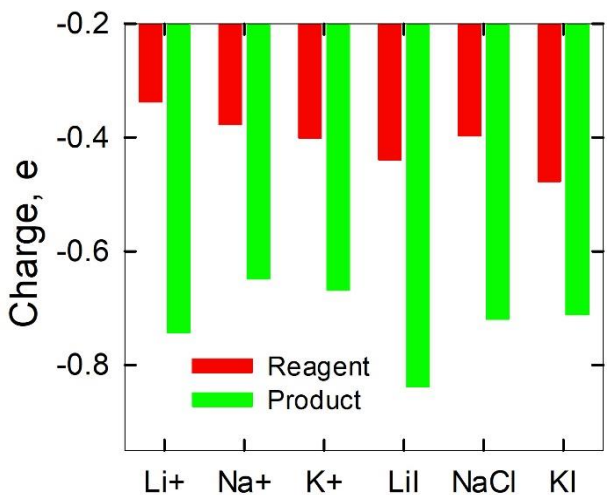

Figure 8. The electrostatic partial charges of the oxygen atom belonging to $CO_2$ (within reactants, red bars) and the carboxyl moiety (within products, green bars) depend on the employed catalyst.

The surface of the electrostatic potential indicates that the most electron-deficient region includes a cation, whereas the electron-richest regions include an anion and oxygen atoms (Figure 9). The hydroxyl proton maintains an H-bond with the chloride atom. The corresponding distance amounts to 214 pm. The samples catalyzed by LiI and KI do not exhibit pronounced hydrogen-halogen patterns due to a larger atomic radius of iodine. For instance, the H…I distance equals 242 pm (with LiI) and 250 pm (with KI). The hydrogen-anion interactions play a paramount role in determining the thermochemistry of deprotonation as we discuss below. Overall, both the reactant state and the product state are dominated by the Coulombic forces based on the partial atomic charges and electrostatic potential surfaces. The high polarity of the reaction medium is decreased at the termination stage of the DMC synthesis, upon passivating the $CH_3OC(O)O^-$ anion with the methyl radical.

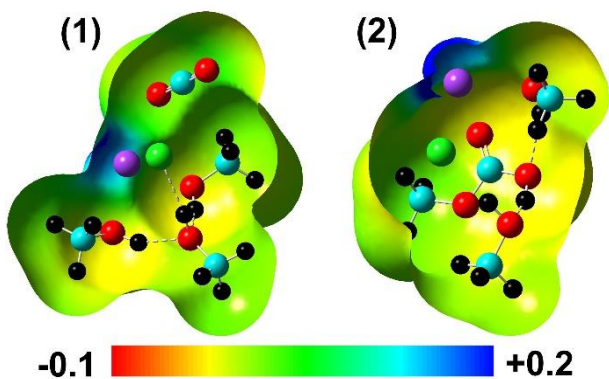

Figure 9. The electron density derived from the total self-consistent-field density (the isovalue equals 4.0×10⁻⁴ a.u.) mapped with the electrostatic potential for (1) reactants and (2) products. The depicted systems contain NaCl as a catalyst particle. The H-bonds are designated by dashed lines.

The reactant molecules strongly attract to the catalyst before the deprotonation of $CH_3OH$. Specifically, the halogen atom of the catalyst coordinates the hydrogen atom of the hydroxyl group of $CH_3OH$. In turn, the alkali atom of the catalyst coordinates the oxygen atom of the hydroxyl group. After the deprotonation, the described coordination patterns are largely retained. Figure 10 depicts localizations of the valence orbitals which are shared by the ions and $CH_3OH$ molecules. The LUMO+1 reveals the partial localization on $K^+$. The valence electron formally belonging to the potassium atom is localized on the oxygen atom of $CH_3OH$. In the LiI-containing system, LUMO (energy of +0.272 eV) is localized on the cation. In the NaCl-containing system, LUMO+4 (energy of +0.318 eV) is localized on the cation. It means that the cations do not have valence electrons in the studied systems. HOMO-2 in the KI-containing sample, HOMO-2 in the LiI sample, and HOMO-3 in the NaCl-containing sample are shared by the iodide/chloride anions and the hydroxyl groups of $CH_3OH$. These shared electron densities reflect the formation of strong electrostatic interactions. In the case of chloride, the designated attractive interaction qualifies for H-bonding. In all systems, the HOMO-4 levels exhibit similar energies and are shared among all $CH_3OH$ molecules. These molecular orbitals corroborate the $CH_3OH$-$CH_3OH$ couplings.

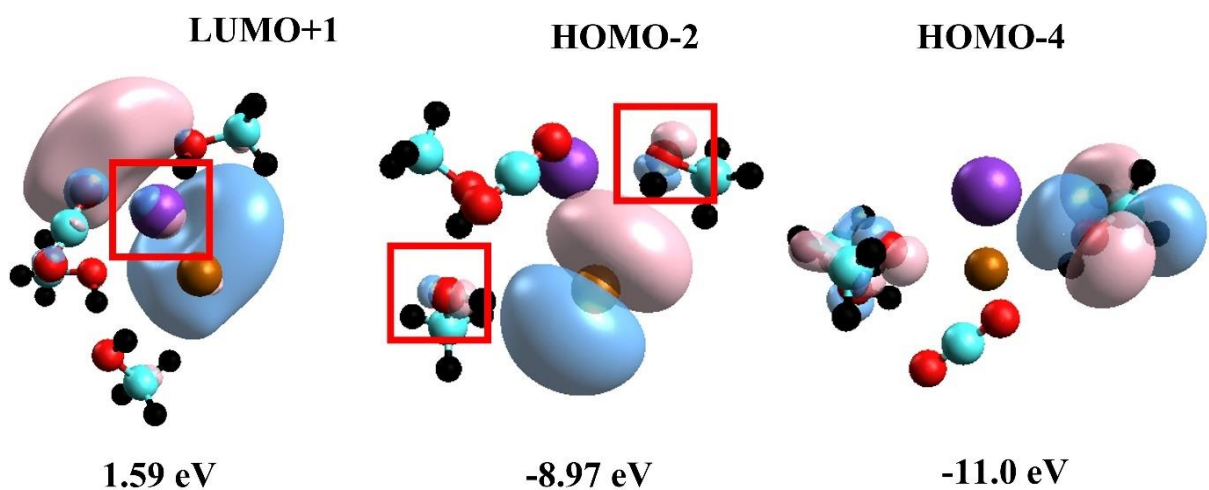

| LUMO+1 | HOMO-2 | HOMO-4 |
|---|---|---|
| 1.59 eV | -8.97 eV | -11.0 eV |

Figure 10. The spatial localization of the selected molecular orbitals in the reactant system containing KI as a catalyst. Potassium is purple, iodide is brown, oxygen is red, carbon is cyan, and hydrogen is black.

### 4.4. Far- and mid-infrared spectra

Figures 11-13 provide infrared spectra in the systems containing LiI, NaCl, and KI. A direct comparison is made between the reactant and product compositions. The depicted C-O band corresponds to symmetric stretching observed in $CH_3OH$ and the carboxyl moiety. C=O represents symmetric stretching in carboxyl, $CO_2$ contributes antisymmetric stretching, and O-H groups manifest as symmetric stretchings in $CH_3OH$ and carboxyl. The reactant's signal at 2404 cm$^{-1}$ corresponds to $CO_2$ and vanishes in the product state (Figure 11).

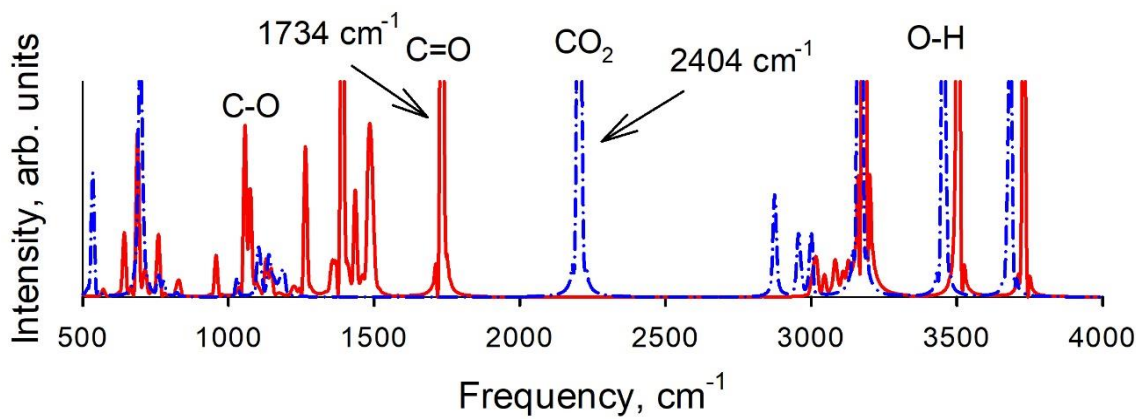

Figure 11. The infrared spectra correspond to the reactant (dashed blue) and product (solid red) of the reaction catalyzed by LiI.

The C-O band corresponds to symmetric stretching in $CH_3OH$ and carboxyl. The signal at 1774 cm$^{-1}$ is symmetric stretching in carboxyl, the C=O double bond. The asymmetric stretching taking place in $CO_2$ shifts by 2 cm$^{-1}$ rightward, as compared to the LiI-containing system. The very strong 162 pm-long hydrogen-oxygen H-bond stretches symmetrically at 2912 cm$^{-1}$. It unites the emerged carbonate moiety with the neighboring hydroxyl group of the solvent. Additionally, vibrational spectroscopy reveals the formation of the O-H…Cl H-bond, 213 pm-long, thanks to a signal at 3451 cm$^{-1}$. This bond exists both within reactants and products. Its length in the product state is 1 pm smaller than in the reactant state, whereas the corresponding frequency in the product state is 3412 cm$^{-1}$ (Figure 12).

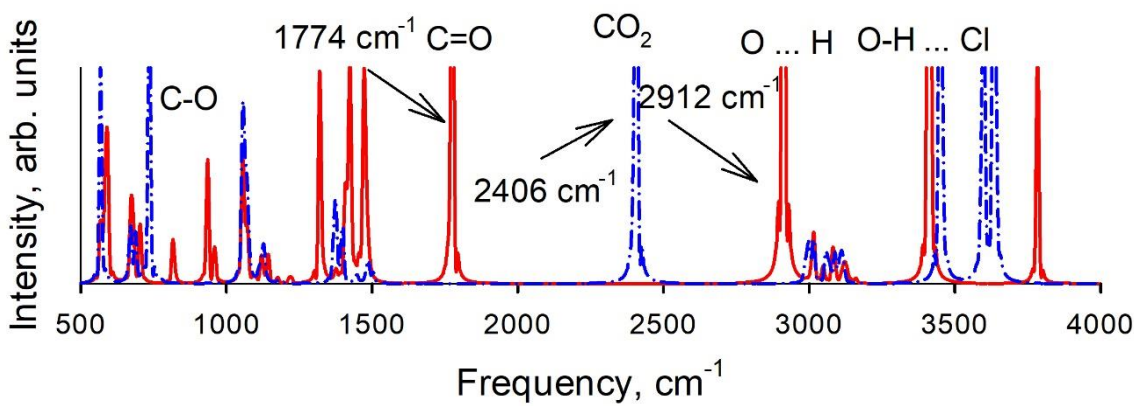

Figure 12. The infrared spectra correspond to the reactant (dashed blue) and product (solid red) of the reaction catalyzed by NaCl.

The KI-catalyzed system exhibits the symmetric C=O vibration within the carboxyl group at 1783 cm$^{-1}$, the antisymmetric stretching of the carbon-oxygen double covalent bond in $CO_2$, and the O-H symmetric stretching band in $CH_3OH$ and -COOH. The key observed frequencies are reasonably similar in all catalyzed samples. This feature suggests the principal similarity of the studied reactions upon the application of the inorganic salts to activate $CH_3OH$ (Figure 13).

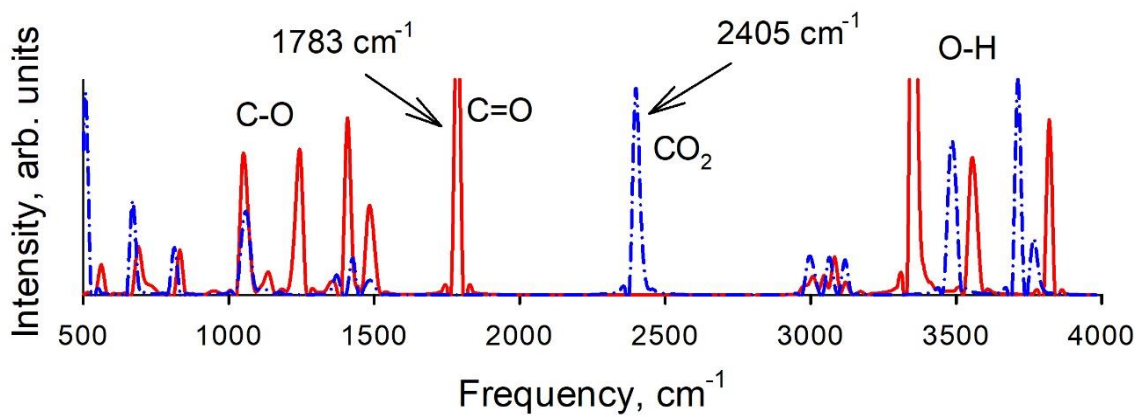

Figure 13. The infrared spectra correspond to the reactant (dashed blue) and product (solid red) of the reaction catalyzed by KI.

### 4.5. Thermochemical potentials

The thermochemical potentials for the deprotonation of $CH_3OH$ and subsequent carboxylation of $CH_3O^*$ are provided in Tables 5-7. The reactions are enthalpically favored and entropically forbidden. The increase in temperature leads to a slight deterioration of the Gibbs free energy. Hence, the yields of the DMC synthesis are expected to be poorer at elevated temperatures. In the meantime, the elevated temperatures are important measures to overcome the activation barriers. The effects of the ion pairs as catalysts at this stage compared to the lone cations are uncertain. For instance, the performance of NaCl, +10.6 kJ/mol at standard conditions, is better than that of $Na^+$, +23.7 kJ/mol. However, the relations are opposite in the cases of the LiI and KI catalysts. In general, the obtained free energies are close to zero suggesting that the discussed stage is a yield-limiting one.

Table 5. The Gibbs free energies of the $CH_3OH+CO_2=CH_3OC(O)OH$ reaction at a set of temperatures and ambient pressure for the various employed catalysts.

| System | Temperature, K | | | | | |
|---|---|---|---|---|---|---|
|  | 250 | 280 | 298.15 | 330 | 360 | 390 |
| $Li^+$ | -2.57 | -1.01 | -0.058 | +1.62 | +3.21 | +4.80 |
| LiI | +23.9 | +25.4 | +26.4 | +28.2 | +29.9 | +31.7 |
| $Na^+$ | +21.3 | +22.8 | +23.7 | +25.4 | +27.0 | +28.7 |
| NaCl | +8.24 | +9.72 | +10.6 | +12.3 | +13.9 | +15.5 |
| K$^+$ | -0.205 | +0.764 | +1.35 | +2.36 | +3.31 | +4.25 |
| KI | +23.6 | +26.5 | +28.2 | +31.4 | +34.4 | +37.6 |

Note that the catalysts should not influence the thermochemistry of the aggregate DMC synthesis. However, they do impact the energetics of the individual reaction stages due to the somewhat different coordinations of the reactant and product species. The locations of the catalyst particles during the reaction may boost or deteriorate the feasibility of certain elementary processes. The temporarily acquired difference in potential energy is returned upon the final stage of the DMC synthesis, which is $CH_3OC(O)OH+CH_3I=CH_3OC(O)OCH_3+HI$. The catalyst ions reorient and alter their potential energies in response to the formation of new molecules. Yet, the effects of the catalysts on the individual stages of the synthesis have a physical sense and contribute to the feasibility of the local chemical transformations.

Table 6. The enthalpies of the $CH_3OH+CO_2=CH_3OC(O)OH$ reaction at a set of temperatures and ambient pressure for the various employed catalysts.

| System | Temperature, K | | | | | |
|---|---|---|---|---|---|---|
| | 250 | 280 | 298.15 | 330 | 360 | 390 |
| Li$^+$ | -15.5 | -15.6 | -15.7 | -15.8 | -15.9 | -15.9 |
| LiI | +11.3 | +10.5 | +10.1 | +9.44 | +8.85 | +8.29 |
| Na$^+$ | +9.08 | +8.50 | +8.16 | +7.60 | +7.09 | +6.61 |
| NaCl | -3.84 | -4.27 | -4.52 | -4.93 | -5.27 | -5.59 |
| K$^+$ | -8.27 | -8.25 | -8.22 | -8.14 | -8.02 | -7.88 |
| KI | +0.646 | -0.333 | -0.903 | -1.87 | -2.74 | -3.57 |

The Li-ion-containing system exhibits the most favorable enthalpy and Gibbs free energy. This phenomenon might have been predicted based on the strong electrostatic attraction of this cation to the polar reactants and products. One should keep in mind that the height of the activation barrier discussed below represents a more important descriptor of the reaction. The NaCl-containing system also displays a competitive thermochemistry. The role of the chloride anion strongly coupling with the hydrogen atom of $CH_3OH$ must be highlighted.

Table 7. The isobaric heat capacities, standard entropies, and entropic factors, $-T\Delta S$, of the $CH_3OH+CO_2=CH_3OC(O)OH$ reaction for the various employed catalysts.

| System | $C_p$, J mol$^{-1}$ K$^{-1}$ | $\Delta S$, J mol$^{-1}$ K$^{-1}$ | $-T\Delta S$, kJ mol$^{-1}$ |
|---|---|---|---|
| Li$^+$ | -3.93 | -52.5 | +15.7 |
| LiI | -23.7 | -54.7 | +16.3 |
| Na$^+$ | -19.1 | -52.2 | +15.6 |
| NaCl | -14.2 | -50.8 | +15.2 |
| K$^+$ | +0.982 | -32.1 | +9.57 |
| KI | -32.2 | -97.7 | +29.1 |

**4.6. Transition states and reaction energy profiles**

We assume that the deprotonation of $CH_3OH$ and the carboxylation of the emerged $CH_3O*$ species take place simultaneously. Figure 14 demonstrates a starting geometry that was used to obtain the second-order saddle point. The conventional covalent hydrogen-oxygen and oxygen-carbon bonds, ~110 pm long and 140 pm long, were stretched forcibly by roughly ~15-20%. The corresponding redundant internal coordinates were added to the Z-matrix. The geometry optimization algorithm was employed to locate the maximum points along the defined degrees of freedom by moving uphill in terms of energy. However, we were unable to find the imaginary vibrational frequency or two frequencies corresponding to the deprotonation and carboxylation within the explored reason. The alternative routes of the reaction were also attempted as follows. First, the proton transfer was attempted to the adjacent $CH_3OH$ molecule. This route was suggested by the initial optimized geometry of the reactant molecular configuration, in which the hydroxyl group of each $CH_3OH$ forms H-bonds with the adjacent solvent molecules. Recall that we represented three $CH_3OH$ molecules explicitly to consider the local reactant structure to the maximum possible extent. Second, the proton transfer was attempted to the chloride/iodide anion. Nonetheless, the movement along this reaction coordinate also did not reveal an imaginary frequency. The subsequent stepwise scanning of both coordinates revealed that the energy smoothly rises upon deprotonation. This observation suggested the absence of the hunted transition state in the direct vicinity of $CH_3OH$.

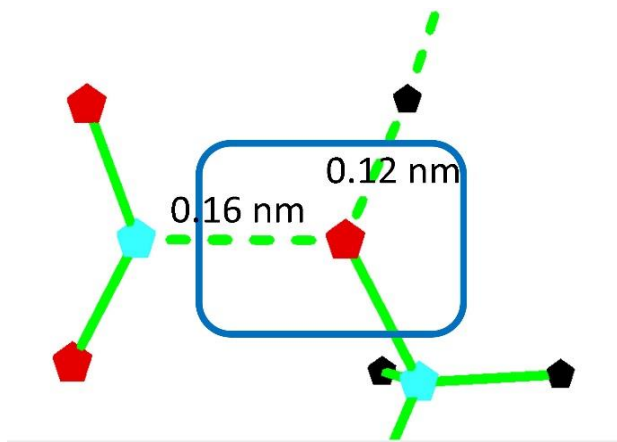

Figure 14. The geometry of the second-order saddle point candidate. The depicted molecular configuration was also used as a starting point to stepwise scan the two reaction coordinates (see dashed lines) by incrementing the bond length by 2 pm. The surrounding $CH_3OH$ solvent molecules were removed for clarity.

To shed light on the energetic alterations at the beginning of the reaction, we compared the thermodynamic stabilities of the proton solvated by liquid $CH_3OH$ to the proton joining the catalyst and also remaining solvated. The energy profile was recorded for the direct proton transfer from the $CH_3OH$ molecule to the HCl molecule. It was found that $Cl^-$ appears an essentially more preferable destination for the proton. The potential energy difference between the considered states amounts to 60 kJ/mol. Nonetheless, both possible locations of the proton constitute stationary points, that is, they can exist for a certain amount of time. The proton exchange process exhibits an activation barrier of 15 kJ/mol. Such barriers are low and are comparable with the steric (physical) obstacles in liquid phases. They are easily overcome at room and elevated temperatures. We hereby conclude that the anion of the catalyst is a dominant host for the proton of $CH_3OH$. Therefore, the halogen-involving reaction pathway must be further explored, whereas the $CH_3OH$-involving reaction pathway may be neglected.

For comprehensiveness of the reaction description, we also investigated the simplified related patterns, one by one. Figures 15-16 provide the reaction energy profiles for the hydrogen abstraction in pure $CH_3OH$ and in the presence of the catalysts. The presence of $CO_2$ in the solution was at this stage deliberately neglected. All catalysts exhibit positive effects on the deprotonation by polarizing the hydroxyl moiety's covalent bond. In the dilute solutions, the strongest positive impact was recorded with $Li^+$, whereas the effects of sodium and potassium were similar. In the cases of the more concentrated solutions, the strongest decrease in the energy cost of the proton abstraction reaction was recorded in the NaCl-containing system.

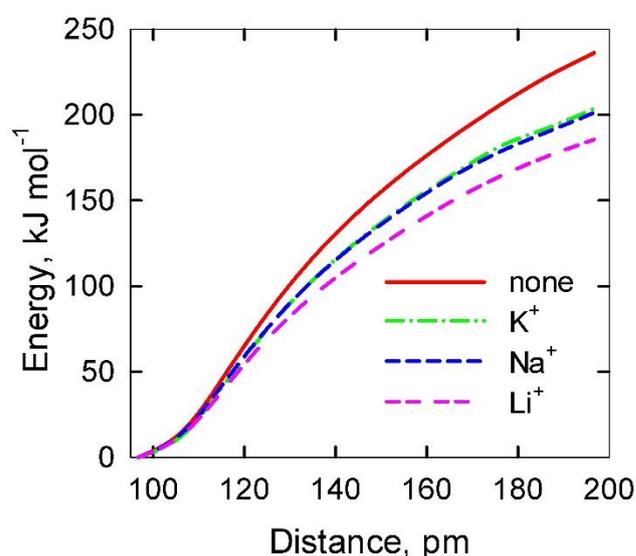

Figure 15. The energy cost of the $CH_3OH$ molecule deprotonation in neat liquid and the presence of the $Li^+$, $Na^+$, and $K^+$ catalysts. $CO_2$ does not participate in the hereby studied deprotonation. See the legend for the designation of lines.

The difference in performance of NaCl relative to LiI and KI can be rationalized through H-bonding between the chloride anion and the hydroxyl hydrogen atom. This strong attraction weakens the bond and simplifies the emergence of the $CH_3O^*$ moiety, which subsequently binds $CO_2$ without an activation barrier. In turn, the electrostatic attraction occurring between iodine and hydrogen is less thermodynamically favorable. According to conventional IUPAC-recommended terminology, only smaller halogen atoms – fluorine and chloride – form H-bonds, whereas larger halogens – bromine and iodine – do not.

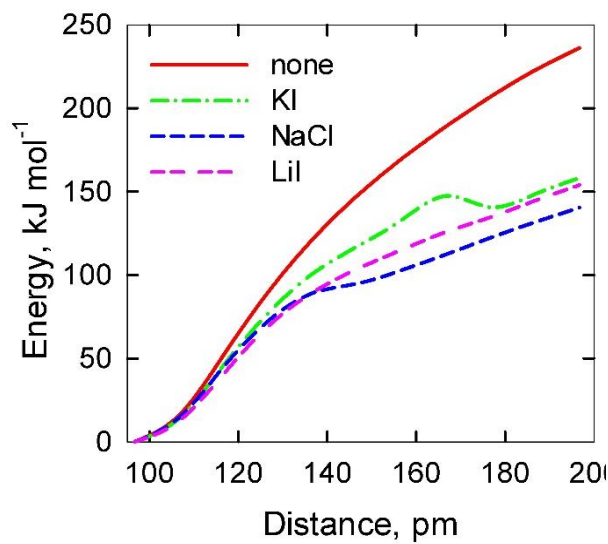

Figure 16. The energy cost of the $CH_3OH$ molecule deprotonation in neat liquid and the presence of the LiI, NaCl, and KI catalysts. $CO_2$ does not participate in the hereby studied deprotonation. See the legend for the designation of lines.

While the inorganic ionic catalysts exhibit definitely positive effects on the proton abstraction, the corresponding energy costs remain high. Furthermore, the penalizing energies smoothly rise without drawing an energy maximum, i.e., activation barriers. Neither the emerging hydrogen halogenides nor $CH_3OH$ solvating the proton is enough thermodynamically favorable to give rise to stationary points. The effects of the alkali metal cations coordinating the $CH_3O^*$ radicals are also not strong enough. The reaction intermediates must be stationary points as a rule. Furthermore, it is clear that the implied activation barriers – to be found – are not commensurate with the experimentally determined yields of DMC. The hereby reported results suggest that $CO_2$ must play an important role from the stage of deprotonation. Its effect cannot be ignored upon the proton abstraction. Figure 17 provides the reaction energy profiles for the carboxylation stage. This instance of simulation deliberately neglects the proton of $CH_3OH$ to separate the energy effects along the different reaction coordinates. The carboxylation partially compensates for the proton detachment penalty. For instance, the stage proceeds without a barrier and brings over -70 kJ/mol to the system. The effect of an alkali metal coordinating $CH_3O^*$ is not essential.

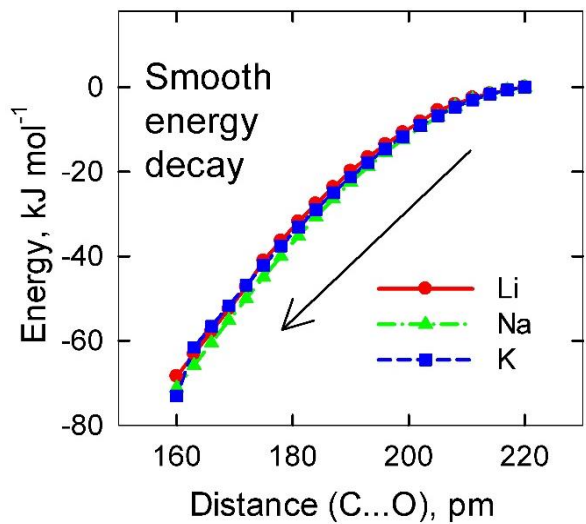

Figure 17. The total electronic energy gains upon $CO_2$ attachment to deprotonated $CH_3OH$ (carboxylation) at various oxygen-hydrogen distances (degrees of deprotonation) in $CH_3OH$. See the legend for the designation of lines.

The joint analysis of the deprotonation and carboxylation returns one to the model, according to which $CO_2$ must form the covalent bond with $CH_3OH$ simultaneously with its deprotonation. Since there is no transition state in the vicinity of $CH_3OH$, it must be located in a different region of the configurational space.

At this stage of research, we changed the reaction coordinates to C-O…H and H…O-C-O, wherein the former designates the distance within the hydroxyl group and the latter designates the distance from the hydroxyl hydrogen atom to the oxygen atom of $CO_2$ (emerging carboxyl group). The O…C reaction coordinate corresponding to the $CO_2$ attachment to $CH_3OH$, which was seen as a natural choice previously, was abandoned. It was possible to obtain a set of first-order saddle points employing such a setup (Table 8). The identified transition states feature high imaginary frequencies, which correspond to antisymmetric radial stretchings. The similar imaginary frequencies reflect similar shapes of the potential energy landscapes in these regions of the potential energy landscapes. The highlighted similarities have been confirmed by the key interatomic distances, such as O…C, H…O-C-O, and H…O-C, see Table 8.

Table 8. The parameters characterizing the high-energy transition states pertaining to hydrogen transfer from the hydroxyl group of $CH_3OH$ to the carboxyl group of methyl hydrogen carbonate.

| Catalyst | Barrier height, kJ/mol | O…C distance, pm | H…O-C-O distance, pm | H…O-C distance, pm | Imaginary frequency, $cm^{-1}$ |
|---|---|---|---|---|---|
| $Li^+$ | 186.8 | 150 | 134 | 119 | 1653i |
| LiI | 186.6 | 146 | 136 | 118 | 1667i |
| $Na^+$ | 172.8 | 147 | 134 | 120 | 1678i |
| NaCl | 188.3 | 146 | 135 | 119 | 1666i |
| $K^+$ | 171.5 | 148 | 134 | 120 | 1678i |
| KI | 178.0 | 150 | 135 | 119 | 1668i |

Figure 18 visualizes the geometries of the two stationary points, $CH_3OH + CO_2$ and $CH_3OC(O)OH$, and the transition state connecting them. Table 8 summarizes all the applicable descriptors. Furthermore, we highlight the locations of the catalyst ions relative to the reactant and product species as well as the activated complex of the reaction. In all considered cases, the NaCl catalyst rigorously coordinates the polarized hydrogen and oxygen atoms. While the coordination in the reactant state activates the deprotonation reaction, the coordination in the product state thermodynamically stabilizes the carboxylated species. These molecular configurations confirm the importance of the inorganic salt catalysts for the newly found reaction mechanism.

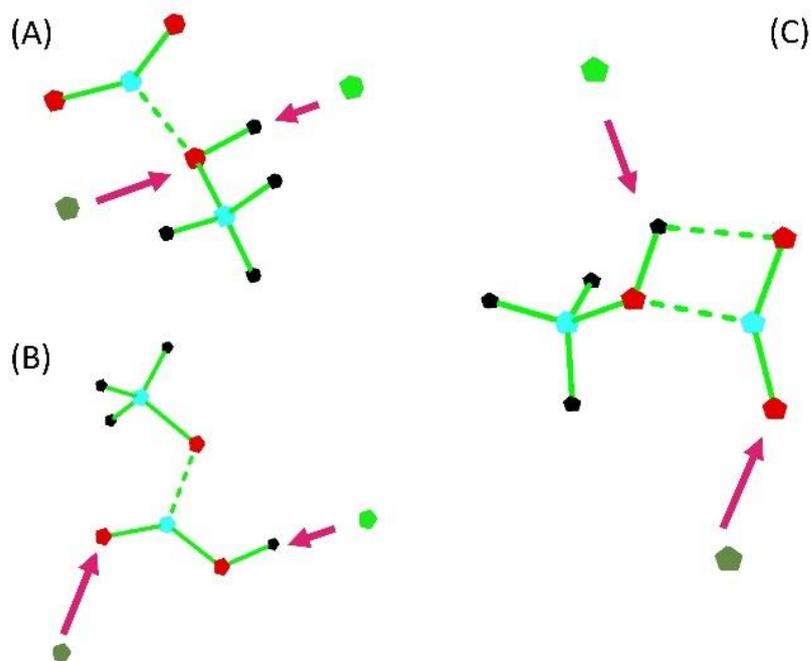

Figure 18. The geometries of the stationary points depict the mutual locations of the reactant/products and catalyst particles at the carbon-oxygen bond length of 160 pm. (A) Beginning of proton transfer. (B) End of proton transfer. (C) Transition state. The lithium atom is dark green and the chlorine atom is light green.

The identified transition states exhibit noteworthy geometries, which point to significant strains in the molecular structures. These structural peculiarities are driven by the geometries of the starting and finishing stationary states since the hydrogen atom must leave $CH_3OH$ at the same time when $CO_2$ acquires its electron and transforms into the carboxyl group. For instance, the intramolecular strain is reflected by an unusually small O…C distance of 146-150 pm. If the atoms were not constrained by the emerging geometry of methyl hydrogen carbonate, the O…C distance must have been about 180 pm. In turn, the H…O breaking bond is restrained less severely, the distances being 118-120 pm. The H…O-C-O distances look most relaxed, 134-136 pm. They correspond to roughly 35% elongation of the forming oxygen-hydrogen covalent bond to finish the building of methyl hydrogen carbonate. The intramolecular strains within the transition state geometries are reflected by the heights of the activation barriers (Table 8). Recall that the barrier height represents an electronic energy difference between the transition state and the reactant state supplemented by the difference in the associated zero-point energies. The identified barriers appear surprisingly high versus our a priori expectations. Based on conventional chemical wisdom coupled with the above-reported ab initio analysis and experimentally obtained DMC yields, see below, we would presently anticipate the major barriers to be lower than 100 kJ/mol.

The activation barriers range from 186.8 kJ/mol in the case of the lone lithium ion to 171.5 kJ/mol in the case of one potassium ion. They are comparable with the energy costs for proton abstraction reported above. Lithium may exhibit a slightly poorer performance because it strongly binds to the carboxyl group and hinders its protonation. Still, the difference between $Li^+$ and $K^+$ is not commensurate with the difference between expected and located activation barriers. Furthermore, we deliberately compared the catalysis by the cation and anion taken together to the catalysis of the cation alone. Lone cations exist in dilute solutions. As the cations in such solutions can catalyze the reaction in the absence of the anion, their effects must be isolated and evaluated properly. Ion pairs exist in relatively concentrated solutions. In such a case, both the cation and the anion coordinate the same $CH_3OH$ molecule. The solutions of high concentrations may contain larger ionic aggregates and they are probably adverse to the catalytic performance.

Thus far, we found that $CO_2$ plays a major role in the determination of the transition state upon the formation of methyl hydrogen carbonate. However, the located transition states have unexpectedly high potential energies due to their strained geometries. While the discussed transition states are formally correct and connect the reactants with the products, they are not likely to populate the lowest-energy reaction pathway. If these barriers were the only ones pertaining to the DMC synthesis reactions, the yield of the reaction must have been way smaller than we know from the experiments. Therefore, the search throughout the potential energy surface for the possible transition states must be continued to find lower-energy barriers. Note that the minimum-energy reaction mechanism is preferentially followed, whereas alternative higher-energy reaction pathways may stay unattended despite formally existing. The goal of the research is to unravel the possibly energy-cheapest route.

Figure 19 demonstrates the newly proposed reaction pathway. First, $CO_2$ collides with the hydroxyl group of $CH_3OH$. The reaction coordinate used to control this stage is C-O…C($O_2$), whose value is stepwise reduced by 1 pm. This phenomenon is associated with a relatively sharp potential energy increase, by over 20 kJ/mol depending on the catalyst, in the reacting system. In response to the emergence of the carbon ($CO_2$)-oxygen ($CH_3OH$) bond, the hydrogen atom of the hydroxyl group elongates its bond with the oxygen atom and forms a stronger H-bond with a halogen anion. After the carboxylation finishes, we rigidly freeze the C-O…C($O_2$) reaction coordinate and activate the H…O-C coordinate. The latter increases stepwise by 1 pm until the transition state is located. The exploration of the energy profile terminates after the carboxyl group of methyl hydrogen carbonate gets protonated. See negative potential energies close to the reaction coordinate value of unity.

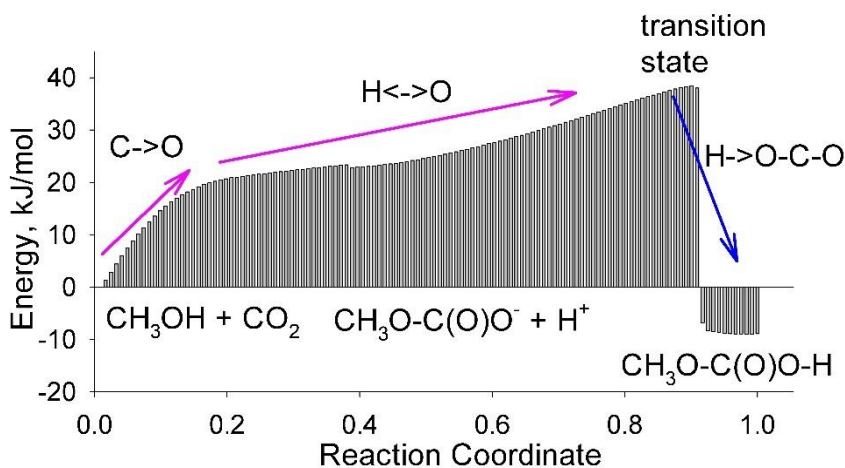

Figure 19. The schematic representation of the minimum-energy reaction path exploration. The depicted reaction coordinate includes a few independent degrees of freedom. The energetic data describing simultaneous deprotonation and carboxylation catalyzed by NaCl was used as a basis for the plot. Note that the plotted numbers are illustrative. The high-quality energies are given in Table 9.

The parameters characterizing the novel set of transition states are given in Table 9. These transition states are substantially different from the previously found set. For instance, the H…O-C distances amount to 197-216 pm as opposed to 118-120 pm in the strained transition state geometries. It is noteworthy that the transition states are located relatively far from the $CH_3OH$ molecule. Based on the interatomic distances, they cannot correspond to the hydroxyl group breakage. Indeed, the investigation of the imaginary vibrational frequencies reveals their association with the rocking vibrations. Note that the proton is covalently bound to the halogen anion rather than to the $CH_3OH$ molecule in these transition states. Consequently, the discovered rocking vibrations involve the halogen anion, the hydrogen ion, and the carboxyl moiety.

Table 9. The parameters characterizing the minimum energy transition states pertaining to hydrogen transfer from the hydroxyl group of $CH_3OH$ to the carboxyl group of methyl hydrogen carbonate.

| Catalyst | Barrier height, kJ/mol | H…O-C distance, pm | H…O-C-O distance, pm | Imaginary frequency, $cm^{-1}$ |
|---|---|---|---|---|
| LiI | 98 | 216 | 189 | 267i |
| KI | 101 | 204 | 232 | 134i |
| NaCl | 38 | 197 | 210 | 150i |
| LiCl | 36 | 201 | 258 | 219i |

The entire revealed reaction mechanism is summarized in Figure 20. The first three frames were recorded upon the O…C($O_2$) reaction coordinate steering, whereas the last three frames were recorded upon pulling hydrogen away from $CH_3OH$. The images highlight the role of the anion in transferring the hydrogen atom from $CH_3OH$ to $CH_3OC(O)O^-$. The formation of the hydrogen-halogen covalent bond greatly decreases the overall energy of the reacting system. Keep in mind that the first stage of steering finished exactly at the value of the O…C($O_2$) reaction coordinate equivalent to the previously identified oxygen-carbon bond length in the methyl hydrogen carbonate. The second stage of steering finished ten steps, i.e., 10 pm after the target molecule formed. As a result, the entire path of the studied reaction was recorded. By definition, this must be the minimum-energy reaction pathway assuming that the chosen reaction coordinates reflect real-world chemistry.

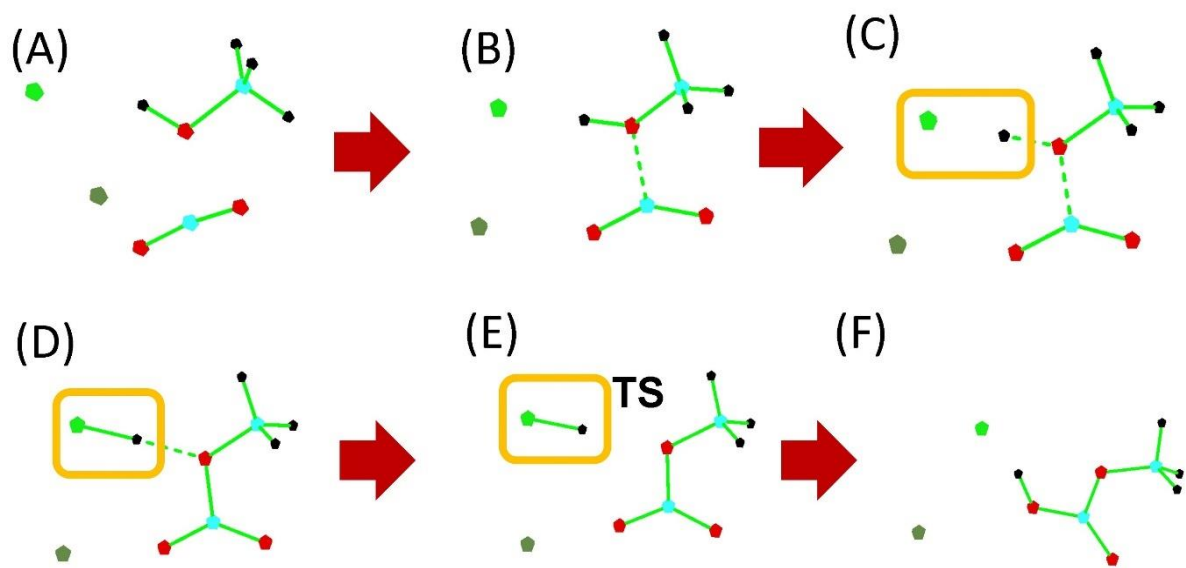

Figure 20. The lowest-energy proton abstraction and simultaneous carboxylation pathway for $CH_3OH$. (A) $CO_2$ approaches $CH_3OH$ upon thermal motion. A cation coordinates hydroxyl oxygen and $CO_2$ oxygen and an anion coordinates hydroxyl hydrogen. (B) $CO_2$ is forming a covalent bond with oxygen. The C…O distance amounts to 170 pm. A cation moves farther from the hydroxyl oxygen. (C) The proton detaches from $CH_3OH$. The C…O emerging covalent bond length is 155 pm. The O…H breaking covalent bond length is 115 pm. (D) The breakage of the O-H covalent bond (at the internuclear separation of 158 pm) and the formation of the H-Cl covalent bond (length of 135 pm). (E) The diffusion of the hydrogen chloride molecule toward the carboxyl group. The H-Cl bond length amounts to 131 pm. (F) The protonation of the carboxyl group. The O-H bond length equals 100 pm. The H…Cl H-bond immediate length equals 153 pm.

Much lower activation energy barriers were hereby located. Relative to the direct scans of the proposed integral reaction coordinate, the nuclei-electronic energies of the stationary points were corrected by the zero-point energies (Table 7). We found that the identity of the anion is greatly important for the catalysis. Compare the barrier of 101 kJ/mol in the system catalyzed by KI to the barrier of 38 kJ/mol catalyzed by NaCl. To confirm the outstanding role of the anion, we computed the activation barrier for LiCl and obtained 36 kJ/mol, which appeared much closer to NaCl than to LiI. The impact of the anion can be rationalized if we consider its role as a carrier of a proton between the two stationary points.

The role of the catalytic action of the cation can also be recorded. For instance, both LiCl and LiI are more successful compared to NaCl and KI. However, the energetic differences are modest. Smaller cations exhibit somewhat better performances thanks to their stronger coordination of the CH$_3$O* radical and saturating CH$_3$OC(O)O$^-$, which represents a part of the activated complex. All the new transition states are much lower in energy, both before and after zero-point energy supplementation than the strained transition states, even though these new complexes are farther from the state of the reactants, in terms of molecular geometry.

Since more trustworthy activated complexes were unraveled and their location along the low-energy reaction pathway was rationalized, we terminated the search at this point. Due to a high energy difference, we expect that only a low-energy line of transition states makes practical sense. The activation barriers of 36-38 kJ/mol are unprecedently small compared to all previous findings in this field. Unfortunately, the yield of DMC is limited by poor thermochemistry, which a low activation barrier cannot compensate for. Yet, faster reaction rates should be expected and the reaction conditions can be made less aggressive. According to the previously published thermochemistry and related energetics,[14,31-33] the pressure elevations somewhat shift the equilibrium of DMC synthesis rightward.

## 5. The experimental synthesis of DMC out of CH$_3$OH using the NaCl catalyst

The goal of the performed experiments is to corroborate the above-discussed theoretical predictions. The NaCl salt was evaluated as a catalyst herein because of the excellent simulation results and the obvious cheapness of this catalyst. The syntheses were carried out under so-called mild conditions. We always look for milder experimental conditions to be more feasible in a real-world industrial process. Milder experimental conditions are less energy-demanding, besides the security and price of the involved equipment. Table 10 summarizes the descriptors of the DMC syntheses by varying the content of the catalyst and temperature in the presence and absence of the dehydrating molecular sieves.

Table 10. The yields and selectivities of the DMC syntheses were obtained by employing various experimental setups. The catalyst is NaCl. The molecular sieves were used for the dehydration of the reaction media. The provided pH values correspond to the end of the reaction.

| # | Sieve (g) | Catalyst (mmol) | T (°C) | Selectivity (%) | Yield (%) | pH |
|---|---|---|---|---|---|---|
| 1 | 2.0 | 5.13 | 80 | 100 | 0.5 | 3.90±0.03 |
| 2 | 2.0 | 8.55 | 80 | 100 | 11.42 | 3.67±0.02 |
| 3 | 2.0 | 11.98 | 80 | 100 | 16.24 | 3.60±0.02 |
| 4 | 2.0 | 15.40 | 80 | 100 | 17.03 | 3.65±0.02 |
| 5 | 2.0 | 18.82 | 80 | 100 | 18.69 | 3.51±0.02 |
| 6 | 2.0 | 25.67 | 80 | 100 | 16.34 | 3.40±0.02 |
| 7 | 2.0 | 34.22 | 80 | 100 | 15.87 | 3.44±0.02 |
| 8 | 2.0 | 11.98 | 90 | 100 | 16.89 | 3.57±0.03 |
| 9 | 2.0 | 11.98 | 100 | 100 | 16.08 | 3.52±0.02 |
| **10** | **0.0** | **11.98** | **80** | **100** | **14.98** | **3.65±0.02** |
| **11** | **0.0** | **18.82** | **80** | **100** | **16.78** | **3.53±0.03** |

The universal parameters: pressure (40 bar); n (CH$_3$OH) = 213 mmol; n (promoter, CH$_3$I) = 20 mmol; reaction duration, t = 24 hours.

The best yields of DMC range from about 17 to about 19%. The usage of extreme CO$_2$ pressures would somewhat increase these values. This is because a higher pressure compensates for an entropic penalty for the binding of the gaseous reactant.[14,31-33] However, the energetic cost of such an improvement would unlikely be sustainable upon technological implementation. The role of the dehydrating agent is marginal (compare entries 3 and 10 to entries 5 and 11) unlike in any competitor technologies. It means that the major route of the DMC synthesis does not produce water. The achievement is unique and greatly simplifies the laboratory setup.

At very small contents of NaCl, the yield of the product is marginal at 80°C and 40 bar. We herein assume that the reaction proceeds too slowly and does not finish in 24 hours. In heterogeneous catalysis, the concentration of the provided catalyst is a cornerstone parameter. The larger quantities of the NaCl catalyst result in higher yields (see entries 1 to 5). However, with the further increase of the catalyst content, the DMC yields deteriorate (see entries 6 and 7). Higher ionic concentrations lead to the formation of large ionic clusters, which lose their catalytic activity.

The effect of the higher temperature of the reactions, 90-100°C, was additionally considered. Since the synthesis of DMC is limited by the high activation barrier (proton abstraction from $CH_3OH$), larger kinetic energies of the reactants may boost the conversion. At the same time, the increase in temperature makes the thermochemistry of the DMC production reaction less favorable, as we reported previously. Therefore, there are two competing factors in action. The reactions conducted at 90°C and 100°C (entries 8 and 9) had little effect on the DMC yield, as compared to entry 3.

The hereby achieved DMC yield is seemingly close to the theoretical limit, which emerges due to the relative thermodynamic stabilities of $CH_3OH$, $CO_2$, and DMC.[14,31-33] This opinion is empirically confirmed and theoretically rationalized by the recent results of Faria and coworkers.[62] Therein, 0.7g of the $CH_3OK$ catalyst was applied in the presence of a molecular sieve under the pressure of 40 bar and the temperature of 80°C. The DMC yield of 17.2% was obtained after 24 hours. The same conditions applied in the present work to NaCl provided a fairly similar DMC yield of 16.24%. NaCl is a much more affordable catalyst than $CH_3OK$. Among other properties, $CH_3OK$ represents quite an unstable compound.

The hypothesis used in the above-reported computer simulations was that the proton that leaves $CH_3OH$ eventually joins one of the available anions, either iodide originating from iodomethane or iodide/chloride originating from the catalyst. At low proton contents, these subproducts must exist in a dissociated state and do not impact the catalytic activity. To verify the identity of the simulated and in-vitro reactions, pH was measured experimentally in the product state (Figure 21). The results indeed indicate an acidic reaction of the media after the reaction. Note that the neutral reaction in $CH_3OH$ corresponds to a pH value of 8.3. These results also suggest that no water is formed during the explored reaction.

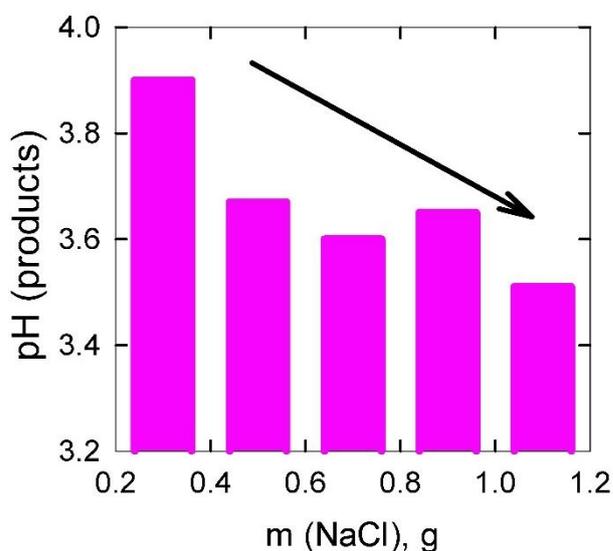

Figure 21. The increase in solution acidity is measured experimentally in the state of the products as higher amounts of NaCl catalysts are applied. The applicable error bars do not exceed 0.03 units.

## 6. Conclusions and final remarks

This work reports a cheap, clean, one-pot, and water-free catalyzed synthesis of DMC. The alkali salts – such as NaCl, KI, and LiI – help $CH_3OH$ to get rid of its hydroxyl proton, which is the most kinetically challenging stage of $CO_2$ valorization. The favorable action of the catalyst is based on the polarization of the reactant's hydroxyl group. The strength of the catalyst is dependent on the activities of the corresponding ions. Smaller cations and anions decrease the energetic cost of the $CH_3OH$ deprotonation more significantly. The performances of NaCl and LiCl appear to be similar, whereas KI and LiI display somewhat weaker catalytic activities. The obtained results suggest that NaCl is the most preferable component of the formulation because this catalyst is the cheapest one.

The achieved yields of DMC are hard-limited by the unfavorable thermochemistry of this reaction.[14,31] The choice of the catalyst cannot alter this descriptor since the catalyst influences only an activation barrier. Alternative reactants must be employed to shift the equilibrium of the reaction rightward. Such reactants must be less thermodynamically stable than the current ones. For instance, a suitable derivative of an alcohol can be applied, such as an alkali metal alcoholate.

We have carried out the most comprehensive exploration of the potential energy landscape for DMC synthesis out of $CH_3OH$ and $CO_2$ thus far. We unraveled two series of transition states, out of which only one is energetically realistic. By separating the integral chemical process into the constituent steps, we proved that the carboxylation of $CH_3OH$ and the deprotonation of $CH_3OH$ take place simultaneously. $CO_2$ acquires the electron from the hydrogen atom of the hydroxyl group and transforms into the carboxyl group as soon as the $CH_3OH$ loses the proton. The protonation of the carboxyl group takes place somewhat later the proton needs to float around the $CH_3OC(O)O^-$ particle following the discovered minimum energy pathway. This novel chemical knowledge is highly instrumental to further screening the catalyst for the present and similar syntheses.

The proposed synthesis of DMC appears much more convenient than the other existing options. First, the current reaction setup is very clean. Second, the reaction takes place in a single phase. Third, the need for a matrix to support solid-state catalysts has been eliminated. Fourth, the main route of synthesis does not produce water excluding the reverse reaction possibility, that is, the decomposition of an obtained product. Fifth, the properties of the reaction hold promise for its smooth extension toward higher alcohols and higher linear carbonates. The hydrocarbon radicals belonging to an alcohol $R_1OH$, and a promoter $R_2I$ can be adjusted. As a result, higher symmetric and asymmetric carbonates can be synthesized, $R_1C(O)OR_2$.

Due to the high demand for organic carbonates and the vivid area of their synthesis, the developed catalytic scheme represents interest to a wide community of organic and physical chemists. Thanks to the demonstrated simplicity in vitro, the method may be robustly commercialized. Furthermore, the synthesis of DMC contributes to the coveted valorization of abundant $CO_2$ and fosters the sustainable development of humanity.

**Acknowledgments**

**Credit Author Statement**

Author 1: Conceptualization; Methodology Development; Validation; Formal analysis; Investigation; Resources; Data Curation; Writing - Original Draft; Writing - Review & Editing; Visualization Preparation; Resources; Supervision; Project administration. Author 2: Investigation; Resources; Visualization Preparation. Author 3: Investigation. Author 4: Formal analysis; Resources; Supervision.

**Conflict of interest**

The authors hereby declare no financial interests and professional connections that might bias the interpretations of the obtained results.

**Author for correspondence**

Inquiries regarding the scientific content of this paper shall be directed to Prof. Chaban at vvchaban@gmail.com.